\documentclass[aps,prl,superscriptaddress,twocolumn]{revtex4}

\usepackage{graphicx} 
\usepackage{amsmath}
 
\newcommand{\beq}{\begin{equation}}
\newcommand{\enq}{\end{equation}}

\begin{document}

\title{Co-existence and shell structures of
several superfluids in trapped three-component Fermi mixtures}
\author{T. Paananen}
\affiliation{Department of Physical Sciences, University of Helsinki, 
PO Box 64, 00014 University of Helsinki,  Finland}
\author{P. T\"{o}rm\"{a}}
\affiliation{Nanoscience Center, Department of Physics,
PO Box 35, 40014 University of Jyv\"{a}skyl\"{a}, Finland}
\author{J.-P. Martikainen}
\affiliation{Department of Physical Sciences, University of Helsinki, 
PO Box 64, 00014 University of Helsinki,  Finland}
\date{\today}

\begin{abstract}
We study the properties of a trapped interacting three component Fermi
gas. We assume that one of the components can have a different mass from
the other two. We calculate the different phases of the three component
mixture and find a rich variety of different phases corresponding to
different pairing channels, and simple ways of tuning the system from one
phase to another. In particular, we predict co-existence of several different
superfluids in the trap, forming a shell structure, and phase transitions
from this mixture of superfluids to a single superfluid when the system parameters
or temperature is varied. Such shell structures realize superfluids with a
non-trivial spatial topology and leave clear observable signatures in the
density profile of the gas.
\end{abstract}

\maketitle

The observation of fermion pairing ~\cite{Regal2004a,Zwierlein2004a,Bartenstein2004b,Kinast2004a,Chin2004a}
and quantized vortices ~\cite{Zwierlein2005a} in ultracold Fermi gases have opened up new
opportunities for exploring
many-body quantum physics. For example, these systems can be used to study also strongly interacting fermionic superfluids
with mismatched Fermi surfaces~\cite{Casalbuoni2004a}. Experiments with such trapped polarized Fermi gases have
revealed exciting and non-trivial phase separation properties~\cite{Zwierlein2006a,Partridge2006a,Zwierlein2006c}.
These system have some intriguing similarities with electronic systems where external magnetic field
imposes unequal mixture of electron spins as well as with quark matter systems where color superconductivity
may appear~\cite{Alford2006a}. In color superconductivity mismatched Fermi surfaces are due to unequal quark masses
and such phenomena might be realized in the core of the compact star or in low-energy heavy-ion collisions.
There is no fundamental problem in trapping and cooling also atoms with different masses
and indeed both boson-boson as well as boson-fermion mixtures 
composed of different atoms have been already experimentally realized~\cite{Roati2002a,Gunter2006a}
and several groups are currently working towards the creation of different fermion-fermion mixtures.
The prospect of degenerate fermion mixtures with unequal masses has naturally 
motivated also recent theoretical contributions~\cite{Petrov2005a,Iskin2006a, He2006a,Wu2006a}.
In this article  we present general phase diagrams for a trapped
three-component mixture of interacting fermions and predict, for instance,
co-existence of several superfluids and superfluid shell-structures.

Some studies related to
three-component Fermi gases have been done~\cite{Honerkamp2004a,Bedaque2006a,Paananen2006a}, but
these studies were restricted either to an unlikely symmetric setup~\cite{Honerkamp2004a},
to equal masses~\cite{Bedaque2006a}, or additional simplifying assumptions were made
concerning the strength of the various couplings in the system~\cite{Bedaque2006a,Paananen2006a}.
Future experiments will certainly be done in an external trap and when the numbers of particles are different, the
trapped physics can be quite different from the physics in a uniform system.
For this reason we will not here only allow different masses and varying coupling strengths in different
pairing channels, but will also solve the realized superfluid states in a trap.
When interactions in all possible pairing channels are included, the number 
of trapped phases can become very high. For example, part of the trap might be occupied
by one type of superfluid, while some other part of the trap is occupied by a different
superfluid or by a normal state. Even though the number of possible configurations at first appears daunting,
using the local density approximation, we find a simple picture which explains 
most of the trapped phases. As it turns out, one can experimentally tune the system from one phase to another, 
by changing either the strength of interactions, 
trapping frequency, atom number, or temperature.

We assume here that the third component has a different mass from the other two components. Consequently,
as a by product, we also solve the trapped physics of the unequal mass two-component Fermi mixtures.
Furthermore, we briefly discuss the strength of induced interaction as the mass ratio
is varied and find that the role of induced interactions becomes more important as the 
mass difference increases.

We employ generalization of the BCS mean-field theory which is expected to be quantitatively fairly reliable
in the weak coupling limit or with stronger interactions at zero temperature. At non-zero
temperatures mean-field theory gives qualitatively reasonable results, 
but many body effects
can alter the critical  temperature substantially. Our Hamiltonian is given by
\begin{eqnarray}
\hat H&=\int d{\bf r}\sum_{\sigma=1,2,3}\left(-\frac{\hbar^2\nabla^2}{2m_\sigma}
+V_\sigma({\bf r})-\mu_{\sigma}\right){\hat 
\psi}^{{\dagger}}_{\sigma}({\bf r}){\hat \psi}_{\sigma}({\bf r})+\nonumber\\
&\Delta_{12} {\hat \psi}^{{\dagger}}_{1}({\bf r}) {\hat \psi}^{{\dagger}}_{2}({\bf r})
+ \Delta^*_{12} {\hat \psi}_{2}({\bf r}) {\hat \psi}_{1}({\bf r})+
\Delta_{13} {\hat \psi}^{{\dagger}}_{3}({\bf r}) {\hat \psi}^{{\dagger}}_{1}({\bf r})\nonumber\\
&+\Delta_{13}^*{\hat \psi}_{1}({\bf r})  {\hat \psi}_{3}({\bf r}) +
\Delta_{23} {\hat \psi}^{{\dagger}}_{3} ({\bf r}) {\hat \psi}^{{\dagger}}_{2}({\bf r}) 
+\Delta_{23}^*{\hat \psi}_{2}({\bf r})  {\hat \psi}_{3}({\bf r}) \nonumber\\
&-|\Delta_{12}|^2/g_{12}-|\Delta_{13}|^2/g_{13}-|\Delta_{23}|^2/g_{32},
\end{eqnarray}
where $\hat \psi^{{\dagger}}_{\sigma}({\bf r})$ and $\hat \psi_{\sigma}({\bf r})$ are fermionic creation 
and annihilation operators of the $\sigma$-component and interaction strengths
are given by $g_{\sigma\sigma'}=2\pi\hbar^2a_{\sigma\sigma'}/\mu_r$, where
$a_{\sigma\sigma'}$ is the scattering length and  $\mu_r=(1/m_\sigma+1/m_{\sigma'})^{-1}$
is the reduced mass of the scattering atoms.For concreteness, most of our calculations are applied to a situations in which
two of the components are different hyperfine states of $^{6}\text{Li}$ atoms
and one of the components consists $^{40}\text{K}$ atoms, although
we also draw some general conclusions with arbitrary mass ratios.
This implies  $m_1=m_2$.
$\Delta_{\sigma\sigma'}({\bf r})=g_{\sigma\sigma'}\langle {\hat \psi}_{\sigma}({\bf r})
{\hat \psi}_{\sigma'}({\bf r})\rangle $ are the order parameters which we find  by minimizing the
grand canonical potential
$\Omega=-k_BT\ln \text{Tr}[e^{-\beta\hat H}].$
Here the $k_B$ is Boltzmann constant, $T$ is the temperature, and $\beta=1/k_BT$.
Unless indicated otherwise, we use the unit of length $1/k_F$, where $k_F$
is the Fermi momentum of the first component ideal gas in the center of the trap. 
Due to the use of contact interaction some integrals in momentum space
diverge and are regularized, as usual, by subtracting 
the divergent contribution away. 
In this article we will focus on the most interesting case 
when all coupling strengths are negative.

Here we take the trapping potentials to be harmonic $V_\sigma({\bf r})=m_\sigma\omega_\sigma^2r^2/2$.
The potentials are identical for the same mass atoms, which is  accurate for the lowest
two hyperfine states of $^{6}{\rm Li}$, for example, but the heavy atoms are assumed to experience
a different trapping potential. This is in general true and having identical trapping potentials
would require extra effort.
In order for the details of our calculations to be valid,
the heavy fermion should nevertheless experience a trap with a same aspect ratio as the light fermions.
Most of our results will nevertheless remain qualitatively valid even with more complicated
trapping scenarios.

In order to compute different configurations we

employ a local density approximation (LDA). In LDA  we can, at each point in space,
use a local chemical potential $\mu_{\sigma}({\bf r})=\mu_{\sigma}-V_\sigma({\bf r})$ and find 
the state which minimizes the grand potential. The chemical potentials in the center
of the trap must then be determined in such a way that atom numbers have desired values.
LDA could also  be used in a more refined way by minimizing energy over the whole system~\cite{Jensen2006a} while including
the possibility of more exotic superfluid phases such as the breach-pairing/Sarma state.

{\it Trapped phases.}---
The minimization of the grand potential reveals that in a uniform system 
one  has four different phases, three different superfluid phases corresponding to different pairing channels
with the remaining component in a normal state, 
and a normal Fermi gas. We find some general rules  in the weak coupling regime which predict
the phase which is energetically favorable at zero temperature. First, pairing will occur only for closely matched Fermi
surfaces. How closely the surfaces must be matched depends on the coupling strengths and stronger coupling
implies greater  tolerance for a Fermi surface mismatch. Second, if several pairing channels have same
coupling strengths the pairing will occur in the channel with involving the
heavy fermion. This effect is due
to the density of states increasing with mass which translates into lower energy. However, it should be kept in mind
that this density of states effect can be masked by increasing the coupling in the channel  involving the
light fermion.

\begin{figure} 
\includegraphics[scale=0.485]{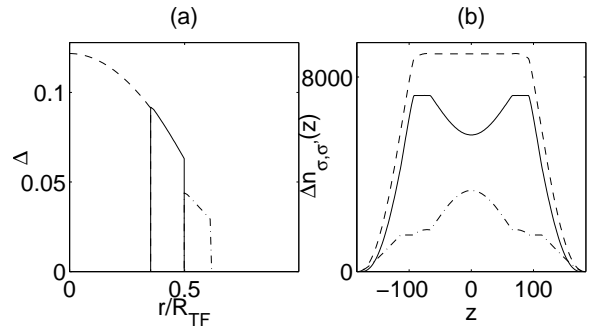} 
\vspace{-0.5cm}
\caption{(a) Gaps as a function of position ($R_{TF}$ is the ideal gas Thomas-Fermi
radius of the second component). Solid, dashed, and dashed-dotted lines describe 
$\Delta_{23}$, $\Delta_{13}$, and $\Delta_{12}$ respectively.
(b) Doubly integrated density differences.
Solid, dashed, and dashed-dotted lines describe $\Delta n_{23}$, $\Delta n_{13}$, and
$\Delta n_{12}$ respectively.
We used the parameters $N_1=6\cdot10^4$, $N_2=5\cdot10^4$, and $N_3=2\cdot10^4$, $k_Fa_{12}=-1.04$,
$(1+1/r_m)k_Fa_{13}=-1.03$, $(1+1/r_m)k_Fa_{23}=-1.15$, where $r_m=m_{Li}/m_{K}$
and $\omega_{3}/\omega_{1}=0.4$.}   
\label{Fig:all_vv}  
\end{figure}

In the trap, the  situation easily becomes more rich.  The variables one can easily vary independently are the
particle numbers of each component $N_{\sigma}$, ratio of the trapping frequencies for
$^{40}\text{K}$ and $^{6}{\rm Li}$ atoms, $T$, and using a (wide) 
Feshbach resonance, at least one of the  coupling strengths.
As we demonstrate in Fig.~\ref{Fig:all_vv} at $T=0$, one can easily find examples of systems with 
all four different phases co-existing in the trap at the same time.
This is made possible by unequal atom numbers and different trapping frequency 
of the $^{40} \text{K}$-component.  

The Fermi surfaces are matched when
local chemical potentials satisfy $\mu_{\sigma}({\bf r})=m_{\sigma}/m_{\sigma'}\mu_{\sigma'}({\bf r})$.
When the  numbers of atoms in the equal mass $12$-channel are different also their chemical potentials in the center of the trap
are different and their Fermi surfaces are not sufficiently matched throughout the system.
For a two component system this would result in  a superfluid core surrounded by 
a normal gas~\cite{Zwierlein2006a,Partridge2006a,Zwierlein2006c,Shin2006a},
but in an interacting three component
system superfluidity in the two unequal mass pairing channels can appear. Since the number of atoms
and trapping frequencies are different, Fermi surfaces can be matched for different pairing channels 
at different locations in the trapped gas. 
The requirement of matched Fermi surfaces implies that it
is most natural to define a dimensionless effective trap frequency
for the third component through
$r_{\omega}=m_{\rm K}\omega_{\rm K}/m_{{\rm Li}}\omega_{{\rm Li}}$.
When $r_{\omega}=1$ the Fermi surfaces which are matched at the origin
remain matched throughout the trap.
For concreteness we assume a harmonic trap for Lithium atoms with a geometric
average of trap frequencies $\omega_{{\rm Li}}=670{\rm Hz}$

Since it is experimentally easier to observe integrated densities as opposed to a three-dimensional
density distributions, in part (b) of Fig.~\ref{Fig:all_vv} we show the doubly integrated density differences 
$\Delta n_{\sigma\sigma'}(z)=\int dxdy n_\sigma({\bf r})-n_{\sigma'}({\bf r})$.
We find that superfluid regions appear as flat areas in the integrated density-differences.
If the number of atoms participating in the superfluidity in some channel is high enough, it can 
leave behind a clearly observable experimental signature in the density distributions, not unlike 
what happens in a polarized two-component systems~\cite{Zwierlein2006c,Shin2006a}.

\begin{figure} 
\vspace{-0.5cm} 
\includegraphics[scale=0.485]{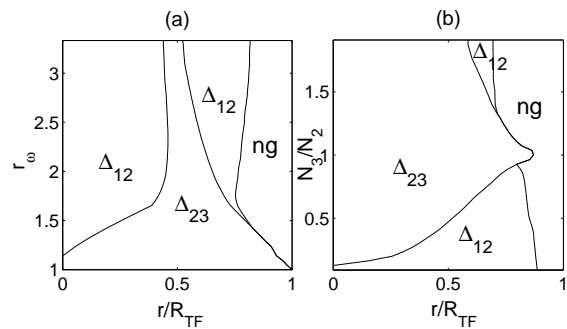}
\vspace{-0.5cm} 
\caption{(a) Zero temperature phase diagram as a function  $r_{\omega}$ and $r/R_{TF}$ with
particle numbers  $N_{1,2,3}=5.5\cdot10^4 $.
(b) Zero temperature phase diagram  as a function
particle number ratio $N_3/N_2$ and $r/R_{TF}$, with $N_1=N_2=5.5\cdot10^4$ and $r_{\omega}=1$.
In both Figures we used  $k_Fa_{12}=-1$, $(1+1/r_m)k_Fa_{23}=-1$, and
$|a_{23}|>|a_{13}|$.N indicates a normal gas or a vacuum at  sufficiently large distances.}
\label{Fig:omegapolar}  
\end{figure}

Let us now  discuss the dependence of the system on experimentally easily tunable parameters.
In Fig.~\ref{Fig:omegapolar} (a) we show the
phase diagram as a function of the effective frequency ratio $r_{\omega}$ and position in a trap. 
For equal  effective trapping the system is a pure $23$-superfluid, due to the
higher density of states for the unequal mass channel.
The area occupied by the $23$-superfluid
becomes smaller with increasing $r_{\omega}$ and contracts to a shell
surrounded by $12$-superfluids. This is caused by the reduction in the area
of sufficiently well matched Fermi surfaces with increasing $r_{\omega}$.

Fig.~\ref{Fig:omegapolar} (b) shows the phase diagram as a function of the particle number ratio $N_3/N_2$.
Because we choose $r_{\omega}=1$, 
$\mu_{\sigma}({\bf r})$ profiles have the same shape for all three components and one can get 
matched Fermi surfaces and superfluidity
in the $23$-channel only if the density of the third component is high enough in the center of the trap. 
For this reason there is no superfluid state $\Delta_{23}$ for small number of $N_3$ atoms. Similar conclusion
applies if the number of atoms in the third component is excessively high.

\begin{figure} 
\includegraphics[scale=0.485]{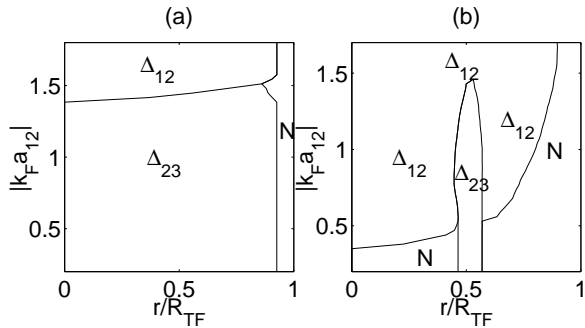}
\vspace{-0.5cm}
\caption{Phase diagrams in the trap as functions $-k_Fa_{12}$ and $r/R_{TF}$ for two different
relative trap frequencies, in (a)   $r_{\omega}=1.0$ and (b) $r_{\omega}=2.67$.
We used  parameters  $N_{1,2,3}=5.5\cdot10^4$, $(1+1/r_m)k_Fa_{23}=1$,
and $|a_{23}|>|a_{13}|$.}
\label{Fig:vvt}  
\end{figure}

In Fig.~\ref{Fig:vvt} (a) we show the phase diagram in the trap as a function $-k_Fa_{12}$.
We find that when one approaches the Feshbach  resonance and  $-k_Fa_{12}$ becomes large there is phase transition
from the superfluid state $\Delta_{23}$ to the superfluid state $\Delta_{12}$. In this case there are  two 
competing two component systems. As the interaction in the same mass channel increases, pairing
in that channel eventually becomes energetically favorable even if the average mass in that channel
is lower. In Fig.~\ref{Fig:vvt} (b) we demonstrate the phase diagram in the trap as a function $-k_Fa_{12}$, when the
trapping potentials are different. 
When $ k_Fa_{12} \in [-0.35,-0.2]$ we find, an unusual configuration,  a normal gas in the center of the trap and a shell
of $23$-superfluid state surrounding it. This can be understood by the fact that pairing with the third component
modifies the density distribution of the components so much that the  Fermi surfaces of the first and second components 
do not match  in the center of the trap anymore. 
A BCS solution is possible, at zero temperature, 
only when $(\mu_1-\mu_2)/2<\Delta_{12}$ and since the gap $\Delta_{12}$ would be small for small  $|k_Fa_{12}|$,
modifications of density distributions do not have to be dramatic to make the BCS solution disappear
in the center of the trap. 

\begin{figure}
\includegraphics[width=0.72\columnwidth]{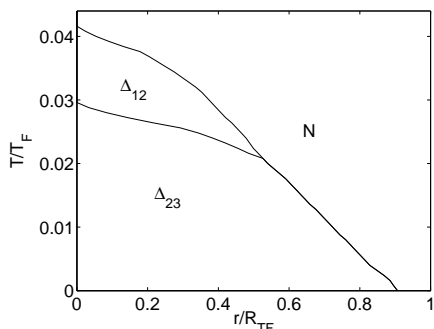}
\vspace{-0.5cm}
\caption[Fig4]{Phase diagram as a function of the temperature $T$ and position.
We used parameters  $N_{1,2,3}=5.5\cdot10^4$, $r_{\omega}=1$, $k_Fa_{12}=-1$, $(1+1/r_m)k_Fa_{23}=-1$, and
$|a_{23}|>|a_{13}|$. With our parameters the Fermi temperature $T_F\approx 600\text{nK}$ and
$R_{TF}\approx 50\, \mu{\rm m}$.
}  
\label{Fig:lampo}
\end{figure}

The mean-field critical temperature for superfluidity between atoms with unequal masses
in the BCS regime can be calculated and is given by
$k_BT_c=[8\gamma\sqrt{r_m}/(e^2\pi)]\epsilon_F
\exp[\pi/(2|k_Fa|)]$, where $r_m=m_{\text{Li}}/m_{\text{K}}$.
Therefore, for the same value of the scattering length the $T_c$
for a system with unequal masses is changed by a factor of $\sqrt{r_m}$ from
the equal mass case. On the other hand at $T=0$ pairing in the unequal mass
channel is energetically favored due to higher density of states of heavy fermions.
Therefore, there is an intermediate (2nd order) transition from the unequal mass
channel to the same mass channel. 
This is demonstrated in Fig. \ref{Fig:lampo}, where we show the phase diagram as of
function of $T$. One clearly sees that as the temperature rises there is first a transition
between the pairing channels, which is then followed by a transition into a normal state. In a trap,
different pairing channels can again coexist in some intermediate temperature window.

Many-body effects due to scattering in a medium modify interactions between fermions and can substantially lower
the critical temperature.
This correction due to induced interactions was calculated by
Gorkov and Melik-Barkhudarov~\cite{Gorkov1961a} and we have generalized this  calculation
to deal with interactions between unequal mass fermions.
We consider interacting atoms on the matched Fermi surfaces with energies equal to the
chemical potentials since in the BCS regime physics close to the Fermi surface dominates.
For the equal mass case the correction in the critical temperature  is the usual $\approx 2.2$, but it increases
as the second component becomes heavier. With $r_m=0.15$ the correction turns out to be $\approx 6.3$. 
Due to this mass dependence of the Gorkov correction, we expect that finite temperature 
mean-field calculations  overestimate
the critical temperature in the channel with unequal masses relative to the equal mass channel.

{\it Conclusions.}---We have solved the mean-field theory of a trapped  three-component Fermi gas 
We found  different superfluids coexisting in a same trapped system which can leave
clear signals into the easily measurable density distributions.Such signatures of
the spatial distribution of the different superfluids could be
complemented by information about the different pairing gaps obtainable by
rf-spectroscopy~\cite{Chin2004a,Kinnunen2004b}.
With the our example parameters the coherence length in the
center of the cloud turns out to be $0.11\cdot R_{TF}$. While this is a fairly large fraction we nevertheless
choose to use the local density approximation. This
is unlikely to affect the general features of our results, which hinge
on pairing with matched Fermi surfaces. By increasing the atom numbers
the local density approximation becomes ever more accurate
In this article we have ignored the surface energy contributions between different superfluids.
However, by applying the results of De Silva and Mueller~\cite{DeSilva2006b}
we have estimated that in our case the surface contribution compared with the
superfluid energy in the shell structures is at worst at the
$2\%$ level and usually much less than this.
Therefore,  they are unlikely to have a large effect on our phase diagrams. Furthermore, the
surface energy compared with the superfluid bulk energy scales in such a way that
it can always be made small by either increasing atom numbers or by reducing the trap frequency.
It would be interesting to understand how co-existing superfluids of
non-trivial spatial topology influence 
the collective and the rotational properties of these systems.

This work was supported by Academy of Finland (project numbers 207083,
106299, 205470) and conducted as part of a EURYI scheme award. See
www.esf.org/euryi.
\vspace{-0.0cm}
\bibliographystyle{apsrev}
\bibliography{bibli}

\end{document}